\documentclass[aps,prc,twocolumn,superscriptaddress,floatfix,amsmath,amssymb,nofootinbib,longbibliography]{revtex4-2}

\usepackage[T1]{fontenc}
\usepackage{url}
\usepackage{physics, mathtools, braket}
\usepackage{soul}
\usepackage{comment}
\usepackage{color}
\usepackage{graphicx}
\usepackage{orcidlink}

\usepackage{hyperref}
\hypersetup{
    colorlinks = true,
    citecolor  = blue,
    linkcolor  = blue,
    urlcolor   = blue
}

\newcommand{\la}{\langle}
\newcommand{\ra}{\rangle}
\newcommand{\Xmax}[1]{\ensuremath{#1_\text{max}}}

\newcommand{\Ethreemax}{\ensuremath{E^{(3)}_\text{max}}}

\newcommand{\cluster}[1]{\ensuremath{\mathcal{T}_{#1}}}

\newcommand{\elem}[2]{{$^{#2}$}\text{#1}}

\newcommand{\ai}{\textit{ab initio}}
\newcommand{\ie}{\textit{i.e.}}
\newcommand{\eg}{\textit{e.g.}}

\newcommand{\defCCSD}{CCSD}
\newcommand{\bogCCSD}{BCCSD}

\newcommand{\defHF}{deformed HF}
\newcommand{\sphHFB}{spherical HFB}

\newcommand{\acroGs}{ground state}
\newcommand{\acroSp}{single-particle}
\newcommand{\acroQp}{quasi-particle}

\newcommand{\magicint}{EM 1.8/2.0}
\newcommand{\deltago}{$\Delta$NNLO$_\text{GO}$(394)}

\newcommand{\MeV}{\text{MeV}}
\newcommand{\fm}{\text{fm}}

\begin{document}

\allowdisplaybreaks

\title{
From closed shells to open shells: Coupled-cluster calculations of atomic nuclei
}

\thanks{This manuscript has been authored in part by UT-Battelle, LLC, under contract DE-AC05-00OR22725 with the US Department of Energy (DOE). The US government retains and the publisher, by accepting the article for publication, acknowledges that the US government retains a nonexclusive, paid-up, irrevocable, worldwide license to publish or reproduce the published form of this manuscript, or allow others to do so, for US government purposes. DOE will provide public access to these results of federally sponsored research in accordance with the DOE Public Access Plan (\url{http://energy.gov/downloads/doe-public-access-plan}).\\} 

\author{F.~Marino \orcidlink{0000-0001-7743-1982}}
\affiliation{Institut f\"{u}r Kernphysik and PRISMA+ Cluster of Excellence, Johannes Gutenberg-Universit\"{a}t Mainz, 55128 Mainz, Germany}

\author{F.~Bonaiti \orcidlink{0000-0002-3926-1609}}
\affiliation{ Facility for Rare Isotope Beams, Michigan State University, East Lansing, MI 48824, USA}
\affiliation{Physics Division, Oak Ridge National Laboratory, Oak Ridge, TN 37831, USA}

\author{P.~Demol \orcidlink{0000-0003-2511-7179}}
\affiliation{Universit\'e Libre de Bruxelles, Institut d’Astronomie et d’Astrophysique, 1050 Brussels, Belgium}
\affiliation{Brussels Laboratory of the Universe -- BLU-ULB, 1050 Brussels, Belgium}
\affiliation{KU Leuven, Instituut voor Kern- en Stralingsfysica, 3001 Leuven, Belgium}

\author{S.~Bacca \orcidlink{0000-0002-9189-9458} }
\affiliation{Institut f\"{u}r Kernphysik and PRISMA+ Cluster of Excellence, Johannes Gutenberg-Universit\"{a}t Mainz, 55128 Mainz, Germany}
\affiliation{Helmholtz-Institut Mainz, Johannes Gutenberg-Universität Mainz, D-55099 Mainz, Germany}

\author{T.~Duguet \orcidlink{0000-0002-7596-3851} }
\affiliation{IRFU, CEA, Universit\'e Paris-Saclay, 91191 Gif-sur-Yvette, France}
\affiliation{KU Leuven, Instituut voor Kern- en Stralingsfysica, 3001 Leuven, Belgium}

\author{G.~Hagen \orcidlink{0000-0001-6019-1687} }
\affiliation{Physics Division, Oak Ridge National Laboratory, Oak Ridge, TN 37831, USA}
\affiliation{Department of Physics and Astronomy, University of Tennessee, Knoxville, TN 37996, USA}

\author{G.~R. Jansen \orcidlink{0000-0003-3558-0968} }
\affiliation{National Center for Computational Sciences, Oak Ridge National Laboratory, Oak Ridge, TN 37831, USA}
\affiliation{Physics Division, Oak Ridge National Laboratory, Oak Ridge, TN 37831, USA}

\author{T.~Papenbrock \orcidlink{0000-0001-8733-2849
} }
\affiliation{Department of Physics and Astronomy, University of Tennessee, Knoxville, TN 37996, USA}
\affiliation{Physics Division, Oak Ridge National Laboratory, Oak Ridge, TN 37831, USA}

\author{A.~Tichai \orcidlink{0000-0002-0618-0685}}
\affiliation{Technische Universit\"at Darmstadt, Department of Physics, 64289 Darmstadt, Germany}
\affiliation{ExtreMe Matter Institute EMMI, GSI Helmholtzzentrum f\"ur Schwerionenforschung GmbH, 64291 Darmstadt, Germany}
\affiliation{Max-Planck-Institut f\"ur Kernphysik, 69117 Heidelberg, Germany}

\begin{abstract}
Coupled-cluster theory is a powerful tool for first-principles calculations of atomic nuclei, enabling accurate predictions of nuclear observables across the Segr\`e chart. 
While coupled-cluster computations are especially efficient at shell closures, extensions have been developed to tackle open-shell nuclei, by exploiting the equation-of-motion method or by expanding the coupled-cluster wave function on top of a symmetry-breaking (either deformed or superfluid) reference state.
In this study, we provide a comprehensive comparison of these different formulations applied to the calcium and nickel isotopes using nuclear two- and three-body interactions from chiral effective field theory.
Based on ground-state energies, two-neutron separation energies, and two-neutron shell gaps, different coupled-cluster computations---based on symmetry-broken reference states and equation-of-motion techniques--- offer consistent descriptions of bulk properties across medium-mass isotopic chains.
\end{abstract}

\maketitle

\section{Introduction}
\label{sec: intro}

The scope of \textit{ab initio} computations has expanded significantly over the last decade, 
thanks to a combination of developments in nuclear interactions, many-body methods, and computational resources~\cite{Hergert2020,Stroberg2021,papenbrock2024}.
These advances have enabled the study of medium-mass nuclei and even selected heavy nuclei, such as $^{208}$Pb~\cite{PbAbInitio,Arthuis:2024mnl,Hebeler22a} and $^{266}$Pb~\cite{Bonaiti:2025bsb}.
Key factors in this progress include new potentials, rooted in the fundamental theory of strong interactions, quantum chromodynamics, via chiral effective field theory ($\chi$EFT)~\cite{MACHLEIDT2024104117,Epelbaum2024},
as well as efficient schemes to include the contribution of three-body forces in large model spaces~\cite{Miyagi2022}.

Methods based on a systematic many-body expansion, such as many-body perturbation theory (MBPT)~\cite{Tichai:2020dna}, in-medium similarity renormalization group (IMSRG)~\cite{Hergert2016Imsrg,Stroberg2021,heinz2025}, self-consistent Green's functions (SCGF)~\cite{Barbieri2004,Soma2020,Soma2020Chiral}, and coupled-cluster (CC) theory~\cite{ShavittBartlett,Bartlett2007,Hagen2014Review,Hagen_2016_rev}, have pushed the limits of \ai{} computations to heavier nuclei, as they provide good accuracy at a moderate computing cost that scales polynomially with the system's size.

Early applications of most of these methods have focused on nuclei with closed shells or closed subshells, where spherical symmetry allows for a drastic reduction of the computational complexity~\cite{hagen2008}. 
However, the majority of atomic nuclei are open-shell, thus calling for extensions.
In general, four strategies have been pursued to tackle open-shell isotopes:
\begin{itemize}
    \item[$i)$] Equation-of-motion (EOM) techniques~\cite{Gour2006,Krylov2008,Hagen2014Review,Jansen2011,Jansen2013}, where the open-shell ground state is constructed as an excitation from the \acroGs{}~of a neighboring closed-shell system.
    \item[$ii)$] Symmetry-breaking techniques~\cite{Soma2011,Signoracci2015,Tichai:2018vjc,Hagen2022,Sun2025}, where a symmetry of the Hamiltonian is dynamically broken in the reference state of an open-shell nucleus, allowing to lift the degeneracy of the spherical orbitals.
    \item[$iii)$] Valence-space methods, where the many-particle problem is divided into a frozen core and a restricted space of active orbitals, such that a final diagonalization is performed within the valence space to capture residual correlations~\cite{bogner2014,jansen2014,Stroberg2019,Tichai:2024cyd}.
    \item [$iv)$] Multi-reference approaches (see e.g.~\cite{Hergert:2014iaa,Tich17NCSM-MCPT,Frosini2021mrI,Frosini2021mrIII,Lyakh2012,Koehn2013}), where a superposition of Slater determinants\footnote{More generally, it can be a linear combination of symmetry-breaking product states, e.g. Bogoliubov states.}  conserving the symmetries of the Hamiltonian is taken as a reference.
\end{itemize}
In this paper, we concentrate on the first two strategies applied in the context of CC theory.

Equation-of-motion techniques have been initially introduced to tackle excited states of closed-shell nuclei within both the CC~\cite{wloch2005,Hagen2014Review} and IMSRG~\cite{Parzuchowski2017,Parzuchowski:2017wcq} frameworks. Within the CC method, they have also proved to be a successful and efficient strategy to access the \acroGs{}~and the low-lying spectrum of nuclei in the vicinity of closed-(sub)shell systems~\cite{Gour2006,hagen2010,Jansen2011,Jansen2013,Ekstrom2014,Marino2025}. Such computations proceed in two steps. First, the similarity-transformed Hamiltonian of the doubly- or semi-magic neighbor is efficiently computed using spherical CC theory. Second, an eigenvalue problem is solved in the space of particle, hole, or particle-hole configurations of the closed (sub)shell, enabling access to the states and nuclei of interest. The flexibility of the EOM-CC ansatz has also been exploited in combination with the Lorentz integral transform technique to address the calculation of response functions in both closed-shell~\cite{Bacca2013,Bacca2014,Miorelli2016,Sobczyk2021,Bonaiti2025monopole} and open-shell nuclei~\cite{Bonaiti2024,Marino2024Nsd, Marino2025}. 

Equation-of-motion CC is an efficient and conceptually simple way of extending the CC scope to open-shell isotopes. 
However, its reach is limited to systems close to magicity. To compute open-shell nuclei, methods that exploit the breaking of symmetries (and their restoration) have been designed and applied with great success over the last 15 years. In such approaches, a single-reference computation is performed, in which the reference state breaks some of the symmetries of the underlying Hamiltonian, enabling one to capture static correlations. This was achieved to tackle superfluidity in singly open-shell nuclei by employing a spherical Hartree-Fock-Bogoliubov (\sphHFB{}) solution violating the U(1) global gauge symmetry associated with particle-number conservation, leading to the computation of Ca and Ni isotopic chains within the framework of Gorkov SCGF~\cite{Soma:2013xha,IS484:2014eke,Soma2020Chiral} and Bogoliubov MBPT~\cite{Tichai:2018vjc}. This strategy was more recently implemented within the realm of CC methods via Bogoliubov CC theory (BCC)~\cite{Signoracci2015}, leading to an \ai{} prediction of the neutron drip line in the tin isotopic chain~\cite{Tichai2024}.

To access doubly open-shell nuclei, it is necessary to perform the CC expansion with respect to a mean-field state that breaks rotational invariance~\cite{Novario2020}, \eg{}, an axially-symmetric deformed Hartree-Fock (HF) wave function. This accurately captures bulk properties of deformed Ne and Mg isotopes~\cite{Novario2020,Sun2025} and of nuclei around $A\approx 80$~\cite{Hu2024Zr,Hu2024Ni}. 
For the computation of rotational bands and transition matrix elements, one needs to restore good symmetries, \eg{}, via angular momentum projection. While the restoration of symmetries is routine for mean field states~\cite{Sheikh:1999yd}, projecting the correlated CC wave function onto good angular momentum has only been addressed recently~\cite{Duguet_2015,Qiu2017}. This yields a quantitative agreement with experimental rotational spectra and B(E2) transition strengths~\cite{Hagen2022,Sun2025,Hu2024Ni}.
The particle-number restoration has been formulated within the BCC framework~\cite{Duguet_2017} and successfully applied to the Richardson Hamiltonian problem~\cite{Qiu2019}, however, it is still lacking an implementation for nuclear \textit{ab initio} calculations.

While restoring symmetries is key for spectroscopic quantities relying on symmetry selection rules, bulk properties like binding energies and charge radii are only mildly affected. The energy gain from angular momentum and particle number projections respectively amount to a few \MeV{} in the Ne and Mg region~\cite{Novario2020,Sun2025} and to less than one MeV in tin isotopes, and tends to decrease for increasing mass numbers~\cite{Hagen2022,papenbrock2024,papenbrock_effective_2022,Tichai2024}.
Consequently, symmetry-broken (unprojected) wave functions offer an adequate and reliable description of bulk properties of medium-mass nuclei.

In this work we benchmark different CC approaches to medium-mass even-even open-shell nuclei.
We apply EOM-CC, BCC, and CC on top of a deformed reference state to compute ground-state energies in the calcium and nickel chains for different chiral interactions. The evolution of ground-state energies, two-neutron separation energies, and two-neutron shell gaps are analyzed along those chains, and the differences between CC frameworks for selected nuclei that can be accessed in all open-shell variants are further investigated.

In Sec.~\ref{sec: CC general}, we review coupled-cluster theory, starting from the single-reference formulation and then addressing the distinctive aspects of symmetry-breaking approaches and EOM-CC. Section~\ref{sec: numerical results} presents the numerical results, while Sec.~\ref{sec: Conclusions} provides conclusions and future perspectives.

\section{Coupled-cluster theory}
\label{sec: CC general}


We want to compute accurate  solutions of the time-independent many-body Schr\"odinger equation
\begin{align}
    H | \Psi_n \ra = E_n | \Psi_n \ra \, .
    \label{eq:seq}
\end{align}
Here, the nuclear Hamiltonian $H\equiv T+V+W$ consists of the (intrinsic) kinetic energy $T$, the two-body  potential $V$ and the three-body potential $W$. We use one-body creation and annihilation operators  $c^\dagger_p$ and  $c_p $, respectively, that fulfill the usual anti-commutation rules for fermions and create single-particle states $|p\rangle = c^\dagger_p|0\rangle$ (where $|0\rangle$ denotes the vacuum). The Hamiltonian reads in second quantization
\begin{subequations}
\label{ham}
\begin{align}
    T  &\equiv \sum_{pq} t_{pq} \, c^\dagger_p c_q \, , \\
    V  &\equiv \frac{1}{4} \sum_{pqrs} v_{pqrs} \, 
    c^\dagger_p c^\dagger_q c_s c_r\, , \\
    W  &\equiv \frac{1}{36} \sum_{pqrstu} w_{pqrstu} \, 
    c^\dagger_p c^\dagger_q c^\dagger_r c_u c_t c_s \, , 
\end{align}
\end{subequations}
where all matrix elements are anti-symmetric and Hermitian. The three-body interaction is approximated as an effective two-body interaction via the normal-ordered two-body (No2B) approximation~\cite{Hagen2007,Hebeler22a,Frosini:2021tuj}. 

Single-reference CC theory~\cite{Bartlett2007,ShavittBartlett,Hagen2014Review,Hagen_2016_rev} provides a systematically improvable ansatz for the nuclear ground state. One starts from a reference state $\ket{\Phi}$ that is a product state (which is not orthogonal to the ground state). Acting  with an exponentiated operator $T$ on it then generates dynamical correlations and yields the correlated \acroGs{}
\begin{align}
    \label{eq: cc gs ansatz}
    | \Psi \ra = e^{ T } | \Phi \ra \, .
\end{align}
Important is here that all excitations contained in the cluster operator $T$ commute with each other. 

There is considerable flexibility in selecting the reference state.
A common choice for $\ket{\Phi}$ is the mean-field solution obtained by solving the self-consistent Hartree-Fock equations,
\begin{align}
    \label{eq: HF determinant}
    | \Phi \ra  \equiv \prod_{i=1}^A c^\dagger_i | 0\ra \, .
\end{align}
For closed-shell nuclei, this reference state preserves rotational symmetry and is therefore an eigenstate of the total angular-momentum operator ($\mathbf{J}^2$) with eigenvalue $J=0$.
For open-shell systems, however, breaking symmetries in the reference state can be beneficial. The implications of starting the CC expansion from a \defHF{} state or from an HFB state that includes pairing correlations are examined in Secs.~\ref{sec:defcc} and~\ref{sec: bogoliubov CC}, respectively.

For the reference state~\eqref{eq: HF determinant} the cluster operator $T$ is given by a sum of $n$-particle-$n$-hole ($n$p-$n$h) excitation operators, 
\begin{align}
    T = \sum_{n=1}^{A} T_{n} \, ,
\end{align}
where the components are defined as
\begin{align}
\label{Top}
    T_n = \frac{1}{(n!)^2} \sum_{a_1 ... a_n i_1 ... i_n} t^{a_1...a_n}_{i_1 ... i_n} 
    c_{a_1}^{\dagger} \cdots c_{a_n}^{\dagger} 
    c_{i_n} \cdots c_{i_1} \,  .
\end{align}
The cluster amplitudes $t^{a_1...a_n}_{i_1 ... i_n}$ are antisymmetric with respect to the permutation among upper/lower indices.  Indices $i_k$ and $a_k$ denote single-particle states that are occupied (holes) and unoccupied (particles) in the reference state, respectively\footnote{The BCC approach, reviewed in Sec.~\ref{sec: bogoliubov CC}, replaces the particle-hole separation with an expansion in terms of quasi-particle states, thus requiring some adaptations of the formalism.
}. In Eqs.~(\ref{ham}), (\ref{eq: HF determinant}), and (\ref{Top}) we used the convention that indices $i,j,k,\ldots$ refer to hole states, that $a,b,c,\ldots$ refer to particle states, and that $p,q,r,\ldots$ refer to any single-particle state. We will follow this convention throughout the paper.   

In principle, an exact wave function is obtained by including all terms up to excitation rank $A$. In practice, the expansion of $T$ is truncated at a certain rank of particle-hole excitations. The most common approximation is coupled-cluster with singles and doubles (CCSD), which consists in keeping terms up to 2p-2h excitations, \ie{}, $T \approx T_1 + T_2$.
Higher accuracy is achieved by including 3p-3h (triples) contributions from the $T_3$ operator through a variety of approximations~\cite{lee1984,Bartlett2024,Hagen2014Review}. 
In the Hartree-Fock basis, triples contributions in the cluster operator typically lower the \acroGs{}~energy by about $10\%$ of the CCSD correlation energy. This estimate, well-established in quantum chemistry~\cite{Bartlett2007}, applies also in nuclear physics for both  CC~\cite{hagen2009b,Novario2020,sun2022,Sun2025} and BCC~\cite{demol2024} calculations. In the present work, the CCSD approximation is employed in all computations. 

We normal-order~\cite{ShavittBartlett,Hebeler22a,papenbrock2024} the nuclear Hamiltonian with respect to the reference state $\ket{\Phi}$, \ie{}, we rewrite the Hamiltonian such that all operators that annihilate the reference state are to the right. This yields the normal-ordered Hamiltonian
\begin{align}
    H_N \equiv H - \la \Phi | H | \Phi \ra \, ,
\end{align}
where
\begin{align}
    E_\text{ref} \equiv \la \Phi | H | \Phi \ra 
\end{align}
is the energy expectation of the reference state. $H_N$ again consists of one-, two-, and three-body terms.  

Inserting Eq.~\eqref{eq: cc gs ansatz} into the Schr\"{o}dinger's equation Eq.~\eqref{eq:seq} yields a set of equations that determine the $T$ amplitudes and the ground-state energy. The exponential parametrization induces a (non-unitary) similarity transformation on the Hamiltonian
\begin{align}
    \label{eq: similarity transformed H}
    \Bar{H} \equiv e^{-T} H_N e^{T} = ( H_N e^{T} )_{C} \ .
\end{align}
Here the subscript $C$ denotes that only connected terms contribute~\cite{ShavittBartlett}. The unknown singles and doubles cluster amplitudes are  solved for by left-decoupling the similarity-transformed Hamiltonian from 1p-1h and 2p-2h excitations, leading to the coupled set of non-linear amplitude equations
\begin{subequations}
\label{T2eq}
\begin{align}
    \mel{ \Phi_{i}^{a} }{ \Bar{H} }{ \Phi } & = 0 \, , \\
    \mel{ \Phi_{ij}^{ab} }{ \Bar{H} }{ \Phi } &= 0 \, .
\end{align}
\end{subequations}
Here, $\ket{ \Phi_{i}^{a} } \equiv c_{a}^{\dagger} c_i \ket{\Phi_0} $, $\ket{ \Phi_{ij}^{ab} } \equiv c_{a}^{\dagger} c_{b}^{\dagger} c_{j} c_{i} \ket{\Phi_0} $. The amplitude equations are solved iteratively from an initial (perturbative) guess. 
The correlation energy is defined as 
\begin{align}
\label{eq: cc energy}
 \Delta E_0 &\equiv E_0 - E_\text{ref} = \bra{\Phi} \Bar{H} \ket{\Phi} \,.
\end{align}
Matrix elements of the similarity-transformed Hamiltonian are evaluated using the Baker–Campbell–Hausdorff expansion, which allows to rewrite $\Bar{H}$ as a series of nested commutators~\cite{ShavittBartlett}. Since all elements of the cluster operator commute with each other, the sequence of nested commutators terminates naturally. 
This is a key advantage of CC theory with respect to many-body methods based on unitary transformations, such as the IMSRG~\cite{Hergert2016Imsrg}.

The power of CC theory relates to its compact way of including high-order many-body correlations from the exponential operator
\begin{align}
    e^{T} = 1 + T_1 + T_2 + ... + \frac{1}{4}T_1^2 T_2^2 + ... \, .
\end{align}
As a consequence, high rank excitations, such as triples or quadruples, are partly captured already at the CCSD level through operator products, such as $T_1T_2$ or $T_2^2$, in contrast to a linear excitation operator as in configuration interaction frameworks~\cite{Bartlett2007}.
In addition, CC theory is based on a manifestly size-extensive many-body expansion due to the connected character of Eq.~\eqref{eq: similarity transformed H} ~\cite{ShavittBartlett,Bartlett2007}. 
This ensures a linear scaling of the many-body error $\delta E_0$ with the system size,
\begin{align}
    \frac{\delta E_0}{A} \sim \text{const}   , 
\end{align}
and hence allows to reliably carry out calculations for a wide range of (nuclear) many-body systems.

\subsection{Coupled-cluster calculations based on a deformed reference}
\label{sec:defcc}

Dynamical correlations related to quadrupole collectivity are crucial in doubly open-shell nuclei~\cite{Scalesi:2024nao,papenbrock2024}. To compute such systems in single-reference CC, a Slater determinant reference state breaking $SO(3)$ rotational symmetry~\cite{Novario2020,Hagen2022,Sun2025,Hu2024Zr} can be employed. The symmetry-breaking reference state displays a non-zero quadrupole moment $\la \Phi | Q_{20} | \Phi \ra$, where  multipole operators are defined according to
\begin{align}
    Q_{\lambda \mu} \equiv \sum_i r^\lambda Y_{\lambda \mu} \, .
\end{align}
The intrinsic quadrupole moment may be converted to a deformation parameter~\cite{bohrmottelson}
\begin{align}
    \beta \equiv \frac{4\pi}{3R^2_0 A} \la \Phi | Q_{20} | \Phi \ra \, .
\end{align}
Here, the empirical radius $R_0 \equiv 1.2 \cdot A^{1/3}\, \fm$ is used.
This allows for a characterization of different shapes in the intrinsic frame: oblate ($\beta < 0$), spherical ($\beta=0$) or prolate ($\beta>0$).

In this work, the reference state is enforced to maintain axial symmetry, \ie{}, to carry the angular momentum projection $M=0$ as a good quantum number, such that $\la Q_{2\pm2} \ra = 0$ is satisfied. As the reference state remains a Slater determinant, the deformed CC approach is identical to the textbook formulation applied to closed-shell systems, except that HF \acroSp{}~basis states do not carry good angular momentum (but only good projection $J_z$). Consequently, the loss of angular-momentum conservation that propagates to the (truncated) CC wave function comes at an increase in computational cost, as angular-momentum coupling techniques (aka $j$-scheme) can no longer be used to reduce the memory footprint of storing many-body operators~\cite{Hagen2014Review,Hagen2010Prc}.
This makes it more expensive to include triples excitations in open-shell systems.
Applications in neutron-rich neon and magnesium isotopes~\cite{Novario2020,Sun2025} have shown that triples play a similar role in bulk properties of collective systems as in their spherical counterparts\footnote{As discussed in the introduction, the more costly angular momentum projection of the CC wave function is not included in this work, as the corresponding ground-state energy correction is expected to be small~\cite{papenbrock2024}.}.

Let us discuss the numerical cost of CCSD computations. We consider a single-particle basis that consists of harmonic-oscillator states up to (and including) energy $(e_\mathrm{max}+3/2)\hbar\omega$, where $\omega$ is the oscillator frequency. Then there are ${\cal N}={\cal O}(e_\mathrm{max}^3)$ single-particle states~\cite{HagenNam} and (for assessing the cost of calculations where angular momentum is kept as a good quantum number) ${\cal N}_j={\cal O}({\cal N}^{2/3})$ single-$j$ shells~\cite{Hagen2010Prc}. As an example, for $e_\mathrm{max}=12$ we have ${\cal N}=1820$ and ${\cal N}_j=182$.

In CCSD, the most expensive contribution to the solution of the Eqs.~(\ref{T2eq}) scales as ${\cal O}(A^2{\cal N}^4)$ and ${\cal O}(A^{4/3}{\cal N}_j^4)= {\cal O}(A^{4/3}{\cal N}^{8/3})$ for the angular-momentum breaking and conserving approaches, respectively~\cite{Hagen2010Prc}. Here, $A$ is the mass number of the computed nucleus. We see that breaking rotational invariance comes at a computational cost.  

\subsection{Bogoliubov coupled-cluster theory}
\label{sec: bogoliubov CC}

The Bogoliubov variant of the single-reference CC ansatz~\cite{Signoracci2015,Tichai2024} is based on a particle-number-breaking Bogoliubov vacuum, 
\begin{align}
    \label{eq: Bogoliubov vacuum}
    | \Phi \ra \equiv \prod_k \beta_k | 0 \ra \, ,
\end{align}
which generalizes Eq.~\eqref{eq: HF determinant}.
The \acroQp{}~operators $\{\beta_k\}$ are defined through a unitary Bogoliubov transformation
\begin{align}
    \beta_k  \equiv \sum_p \left(U^*_{pk} \, c_p + V^*_{pk} \, c^\dagger_p \right) \, , \\
    \beta^\dagger_k  \equiv \sum_p\left( U_{pk} \, c^\dagger_p + V_{pk} \, c_p\right) \, ,
\end{align}
where the transformation matrices $U$ and $V$ are obtained from a self-consistent solution of the spherical HFB equations. Due to the particle-number-breaking character of the reference state, the Hamiltonian is replaced by the grand potential
\begin{align}
    \Omega \equiv H -  \lambda_N N - \lambda_Z Z \, ,
\end{align}
where the neutron (proton) chemical potentials $\lambda_N$ ($\lambda_Z$) are used to ensure the correct average neutron (proton) number.

Within the BCC framework, it is convenient to express any operator $O$ in the \acroQp{}~basis via the application of Wick's theorem with respect to the Bogoliubov vacuum Eq.~\eqref{eq: Bogoliubov vacuum}~\cite{Signoracci2015,Duguet_2015}, \ie{}, $O$ is expressed as a sum of terms of the form, \eg{},
\begin{align}
    O^{[31]} \equiv \frac{1}{3!1!} \sum_{k_1 k_2 k_3 k_4} \bar o^{[31]}_{k_1 k_2 k_3 k_4} \, \beta^\dagger_{k_1} \beta^\dagger_{k_2} \beta^\dagger_{k_3} \beta_{k_4} ,
\end{align}
where the superscript $[31]$ indicates the number of \acroQp{}~creation and annihilation operators. Similarly, the cluster operators are represented in the \acroQp{}~basis as
\begin{align}
    \cluster{n} \equiv \frac{1}{(2n)!} \sum_{k_1 ... k_{2n}} t_{k_1 ... k_{2n}} \,
    \beta^\dagger_{k_{1}} \cdots \beta^\dagger_{k_{2n}} \, ,
\end{align}
involving only excitation components\footnote{The notation $\cluster{}$ is introduced to distinguish the quasi-particle variant of the cluster operator from its traditional counterpart.}.
Similar to standard CC theory, the similarity-transformed grand-canonical potential is introduced via
\begin{align}
    \bar \Omega \equiv e^{-\cluster{}} \, \Omega \, e^\cluster{} \, ,
\end{align}
which enters the BCC amplitude equations,
\begin{subequations}
\label{bccamps}
    \begin{align}
        \la \Phi^{k_1 k_2} | \bar \Omega | \Phi \ra &= 0 \, , \\
        \la \Phi^{k_1 k_2 k_3 k_4} | \bar \Omega | \Phi \ra &= 0 \, ,
    \end{align}
\end{subequations}
that decouple the HFB reference state from elementary two and four \acroQp{}~excitations $| \Phi^{k_1 k_2} \rangle \equiv \beta^{\dagger}_{k_1}\beta^{\dagger}_{k_2}| \Phi \ra$ and $| \Phi^{k_1 k_2 k_3 k_4}\rangle  \equiv \beta^{\dagger}_{k_1}\beta^{\dagger}_{k_2}\beta^{\dagger}_{k_3}\beta^{\dagger}_{k_4}| \Phi \ra$, respectively. 
Following the standard CC convention, the approximation $\cluster{}\approx \cluster{1}+\cluster{2}$ is labeled as Bogoliubov CC with singles and doubles (BCCSD). Presently, rotational invariance is enforced in large-scale BCC implementations~\cite{Tichai2024}.

The absence of a particle-hole characterization of \acroSp{}~states in BCC requires the storage of the cluster operator in the full \acroSp{}~basis. This comes at a considerable increase in computational resources.
As there is no distinction between particles and holes, the computational cost to solve the BCCSD equations~(\ref{bccamps}) scales as ${\cal O}({\cal N}_j^6)={\cal O}({\cal N}^4)$~\cite{Signoracci2015}. The comparison with the estimates presented at the end of Sect.~\ref{sec:defcc} shows that spherical BCCSD is more expensive than spherical CCSD and a factor $A^2$ less expensive than CCSD based on an axially symmetric reference state. Breaking both particle number and angular momentum would scale as ${\cal O}({\cal N}^6)$.

In addition, it is crucial to monitor the breaking of particle number by tracking corrections to the particle-number expectation value on the BCC ground state,
\begin{align}
    \Delta A = \la \Phi | \bar A | \Phi \ra \, .
\end{align}
where $\bar{A} \equiv e^{-\cluster{}} \, A \, e^\cluster{}$ is the similarity-transformed number operator.  
This may induce significant particle-number shifts and hence strong contamination of nuclear ground-state energies as $\Delta E \sim \Delta A \cdot 8 \, \MeV$. To this end, a self-consistent procedure is used to constrain the average particle number to its physical value using an additional micro iteration within each BCC loop to determine the amplitudes~\cite{Tichai2024,demol2024}.


\subsection{Equation-of-motion approach}
\label{sec: eom theory}

The EOM-CC approach provides a pathway to investigate open-shell nuclei in the vicinity of closed-shell ones without breaking symmetries in the many-body wave function~\cite{Gour2006,Jansen2011,Jansen2013,Piecuch2013,Musial2003,Bonaiti2024,Marino2025}.

An EOM computation consists of two steps.
First, a closed-(sub)shell nucleus of mass number $A^{*}$ is identified in the vicinity of the target nucleus, and its \acroGs{}~wave function $\ket{\Psi}$ is determined by solving the standard spherical CC equations for the $T$ amplitudes, yielding a \acroGs{}~correlation energy $\Delta E_0^{*}$.
In a second step, the target nucleus with mass number $A = A^{*} \pm k$ ($k =1, 2$) is interpreted as a generalized excitation of the closed-shell system. The present work focuses on $k = 2$, and more specifically, we detail the two-particle-removed (2PR) variant~\cite{Jansen2011,Marino2025,Piecuch2013} in which the target nucleus differs by the removal of two nucleons from the (sub)shell closure.
States $\ket{\Psi_f^{(A^{*}-2)}}$ of a 2PR open-shell system are parametrized by acting on the correlated wave function of its closed-shell neighbor with the excitation operator $R_{f}^{(A^{*}-2)}$ according to~\cite{Jansen2011,Marino2025}
\begin{align}
    \label{eq: eom 2pr right}
    \ket{\Psi_f^{(A^{*}-2)}} \equiv R_{f}^{(A^{*}-2)} \ket{\Psi} = R_{f}^{(A^{*}-2)} e^{T} \ket{\Phi}, 
\end{align}
with
\begin{align}
    R_{f}^{(A^{*}-2)} \equiv 
    \frac{1}{2} \sum_{ij} r_{ij} c_j c_i + \frac{1}{6} \sum_{ijka} r_{ijk}^{a} c_a^\dagger c_k c_j c_i +\dots.
\end{align}
In this work, 0p-2h and 1p-3h configurations are included in the EOM excitation operator.
By inserting Eq.~\eqref{eq: eom 2pr right} into the Schr\"{o}dinger's equation $H_N \ket{\Psi_f^{(A^{*}-2)}} = \Delta E_{f}^{(A^{*}-2)} \ket{\Psi_f^{(A^{*}-2)}}$, the eigenenergies of the target nucleus and the unknown amplitudes $r_{ij}$, $r_{ijk}^{a}$ can be determined as the solution to the eigenvalue problem~\cite{Jansen2011,Marino2025}
\begin{align}
    \label{eq: right eom eigval}
    \left( \Bar{H} R^{(A^{*} - 2)}_{f} \right)_{C} \ket{ \Phi } =
    \omega_{f}^{(A^{*} - 2)} R^{(A^{*} - 2)}_{f} \ket{ \Phi },
\end{align}
where the excitation energies of the open-shell nucleus with respect to the closed-shell reference are given by
\begin{align}
\omega_f^{(A^{*} - 2)} = \Delta E_f^{(A^{*} - 2)} - \Delta E_0^{*} \, .
\end{align}
To solve Eq.~\eqref{eq: right eom eigval} one uses a complex Arnoldi algorithm to obtain the first few low-lying eigenstates~\cite{Hagen2014Review}.

Completely analogous is the two-particle-attached (2PA) approach~\cite{Jansen2011,Jansen2013,Bonaiti2024}, where the $A^{*}+2$ open-shell nucleus is described in terms of excitation operators involving 2p-0h and 3p-1h configurations,
\begin{align}
    \label{eq: eom 2pa right}
    R_{f}^{(A^{*}+2)} \equiv
    \frac{1}{2} \sum_{ab} r^{ab} c_a^\dagger c_b^\dagger 
    + \frac{1}{6} \sum_{abci} r_{i}^{abc} c_a^\dagger c_b^\dagger c_c^\dagger c_i.
\end{align}
The 1p-3h (3p-1h) approximation scheme for 2PR-EOM (2PA-EOM) nuclei is accurate for states which are relatively simple in structure with respect to the reference state~\cite{Jansen2013,gulania2021equation,Bonaiti2024,Marino2025}.  More complex states may not be a good target for the particle-removed/attached ansatz.

Let us also discuss the cost of the 2PR and 2PA computations. Working in a angular-momentum preserving scheme, the dimensions of the eigenvalue problems (\ref{eq: right eom eigval}) and (\ref{eq: eom 2pa right}) scale as ${\cal O}(A^{2}{\cal N}^{2/3})$ and ${\cal O}(A^{2/3}{\cal N}^{2})$, respectively.

\section{Results}
\label{sec: numerical results}

\subsection{Calculation details}
\label{sec: calc_details}

We use two nuclear interactions from chiral effective field theory, namely the \magicint{} potential from Ref.~\cite{MagicInteraction} and the $\Delta$-full \deltago{} potential from Ref.~\cite{DeltaGo2020}. Both produce fairly accurate nuclear ground-state energies up to the Sn chain~\cite{Tichai2024}.
Ground-state energies of even-even Ca ($Z=20$) and Ni ($Z=28$) isotopes allow us to gauge the performance of the three CC variants under consideration.

All calculations employ the spherical harmonic oscillator basis for the one-body Hilbert space. The model space includes a total of $13$  major oscillator shells, i.e. $\Xmax{e}=12$.
We choose $\hbar\omega = 16 \, \rm{MeV}$ as the oscillator frequency.  
To keep the number of three-body  matrix elements tractable, an additional cut is imposed by including only configurations that satisfy $e_1+e_2+e_3 \le \Ethreemax$~\cite{Tichai2024} where $e_p$ denotes the energy of particle $p$ in the harmonic oscillator (neglecting zero-point energies).
For the calcium chain, calculations employ $\Ethreemax = 16$~\cite{DeltaGo2020} whereas for the heavier nickel chain $\Ethreemax = 24$ is used~\cite{Tichai2024,Miyagi2022}.
Three-body forces are treated within the NO2B approximation~\cite{Hagen2007,roth2012,Hebeler22a,Frosini:2021tuj,rothman2025}, where an effective two-body  operator is obtained by contracting three-body  matrix elements with a symmetry-conserving one-body density matrix $\rho$ and adding it to the original two-body  matrix elements, \ie{},
\begin{align}
    \tilde v_{pqrs} \equiv v_{pqrs} + \sum_{tu} w_{pqtrsu} \, \rho_{tu} \, .
\end{align}
The explicit form of $\rho$ depends on the specific CC variant and is chosen as a spherical Hartree-Fock density (for closed-shell CC), a spherical fractional-filled Hartree-Fock density~\cite{Frosini:2021tuj,Hu2024Zr,Sun2025Odd} (for CC on top of a deformed reference), or a spherical HFB density (for BCC).
As discussed extensively in Refs.~\cite{Frosini:2021tuj,Hebeler22a}, this ensures a symmetry-conserving nuclear Hamiltonian in the NO2B approximation.

Coupled-cluster computations are performed on top of a deformed mean-field state while BCC calculations rely on a spherical HFB state, both employing the singles and doubles truncation level. The corresponding results are denoted as \defCCSD{} and \bogCCSD{}, respectively.
For EOM calculations, we perform a spherical CCSD computation based on the closed-shell reference from spherical Hartree-Fock and then include up to 3p-1h (1p-3h) configurations in the excitation operators  in  2PA (2PR) open-shell nuclei~\cite{Marino2025,Jansen2013}. This  is denoted as 2PA-EOM (2PR-EOM).

In the following, theoretical uncertainties are assigned to ground-state energies based on the residual model-space dependence of the results and the truncation of the many-body expansion. These are shown as gray bands in Figs.~\ref{fig:calciumenergies},~\ref{fig:nickelenergies},~and \ref{fig:detailed}.
Doing so, a conservative 2\% symmetric error bar is attributed to the correlation energy due to the truncation of the model-space~\cite{Tichai2024}.
The uncertainty arising from omitted triple excitations materializes in an asymmetric band corresponding to a lowering of the CC energies. The magnitude of the band is equal to 10\% of the \bogCCSD{} correlation energy and is centered on the mean of the \bogCCSD{} and \defCCSD{} calculations. As shown in Ref.~\cite{Marino2025}, this turns out to be a reasonable estimate for the many-body uncertainty also for  2PA-EOM (2PR-EOM)  energies when including triples in the closed-shell reference while keeping the EOM truncation scheme at the level of 3p-1h (1p-3h) excitations. 
We note that the uncertainty from the interaction is not accounted for in our theoretical error bands.

\subsection{Ground-state energies}
\label{sec: calcium chain}
\begin{figure*}[t]
    \centering
    \includegraphics[width=\textwidth]{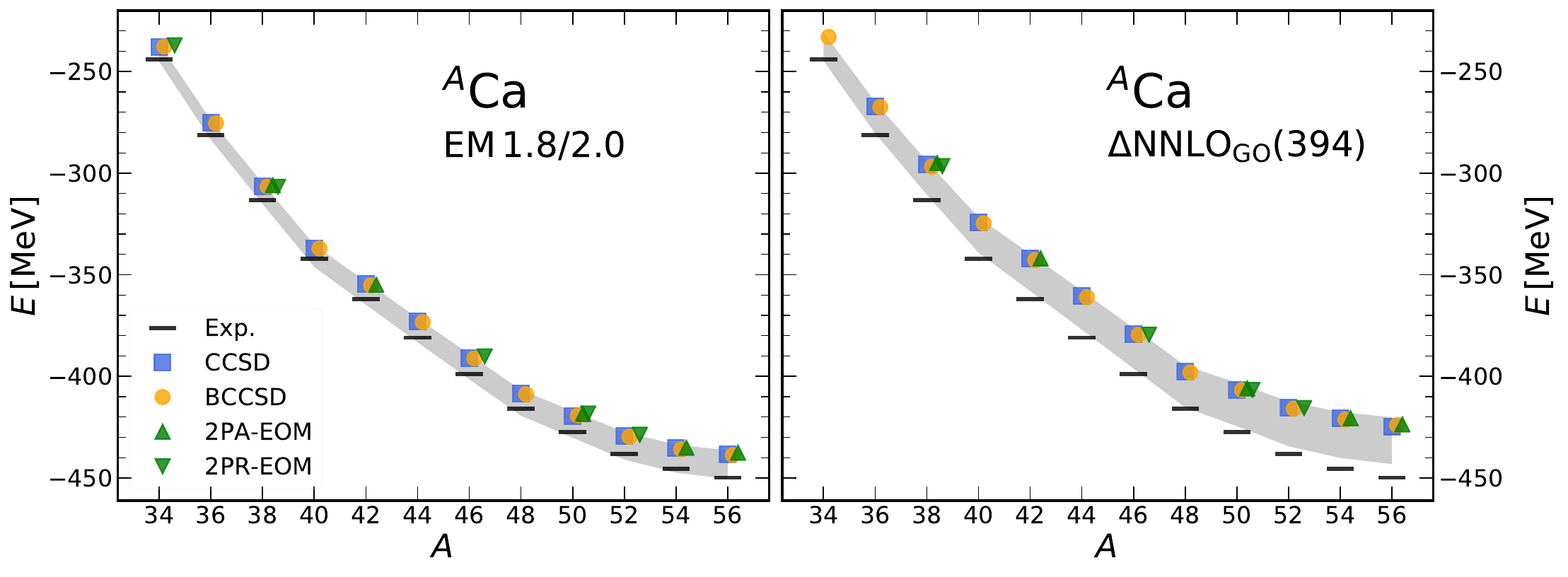}
    \caption{
    Ground-state energies of even Ca isotopes as a function of the mass number $A$.
    The left and right panels display results obtained using the EM 1.8/2.0~\cite{MagicInteraction} and \deltago{}~\cite{DeltaGo2020}  potentials, respectively.
    Calculations performed with \defCCSD{} (squares), \bogCCSD{} (circles), 2PA-EOM (upward triangles), and 2PR-EOM (downward triangles) are reported.
    The gray bands refer to an estimate of the theoretical uncertainties stemming from the model-space and many-body truncations (see text).
    Experimental energies (black) are taken from Ref.~\cite{wang2021ame}.
    }
    \label{fig:calciumenergies}
\end{figure*}

We first compare the different CC results for the ground-state energies of even-even calcium and nickel isotopes. Figure~\ref{fig:calciumenergies} shows \acroGs{}~energies of \elem{Ca}{34-56} obtained from the \magicint{} and \deltago{} interactions.
Nuclei with closed-(sub)shells are \elem{Ca}{36,40,48,52,56} and this allows us to use EOM-CC for neighboring isotopes.

\begin{figure*}[t]
    \centering
    \includegraphics[width=\textwidth]{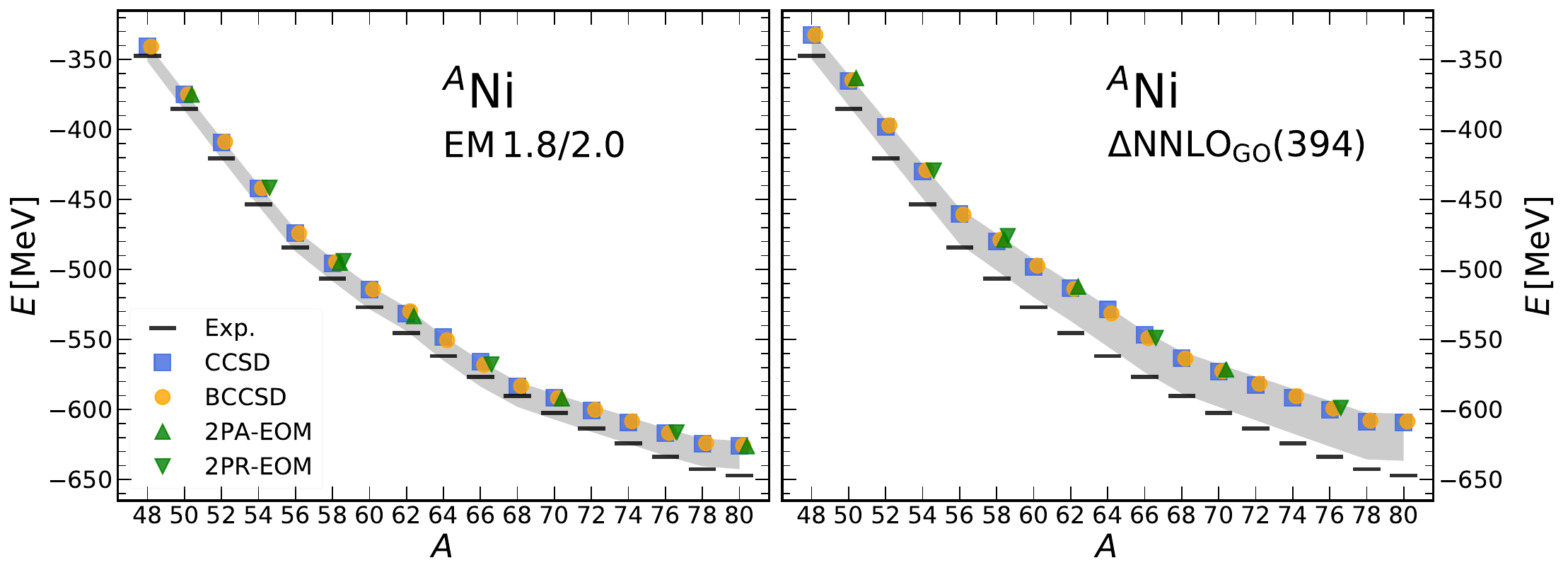}
    \caption{
    Same as Fig.~\ref{fig:calciumenergies} but for Ni isotopes.
    }
    \label{fig:nickelenergies}
\end{figure*}

For both interactions, the different CC methods are in excellent agreement for all nuclei, with their spread being much smaller than the estimated uncertainty.
In general, the CCSD results for the \magicint{} potential are closer to experimental data than \deltago{} because the former interaction is softer than the latter. This is in agreement with previous works~\cite{Hagen2016,simonis2017,Morris2018,Stroberg2019}.
The remaining discrepancy to experimental data is dominated by the lack of triples contributions in the cluster operator, corresponding to about $20 \, \MeV$ in \elem{Ca}{48} for the \deltago{} interaction~\cite{Marino2025}. When the shown triples estimates are included, ground-state energies for the \deltago{} potential are also close to data.

The calcium isotopes show that bulk properties can be well approximated from both deformed and particle-number-broken reference states.
Due to the proton shell closure at $Z=20$, \defHF{} simulations exhibit only small deformation parameters of about $\beta \lesssim 0.1$, and their intrinsic \acroGs{}~configurations are (almost) spherical. 
Therefore, enforcing rotational invariance in a \sphHFB{} calculation may adequately capture the static correlations in the systems. The corresponding pairing energies are of the order of $5 \, \MeV$ with a particle-number variance close to its (nucleus-dependent) minimum value, see Ref.~\cite{Duguet2020}.
As a consequence, both rotational SO(3) symmetry and U(1) global gauge symmetry are only weakly broken, leaving the many-body practitioner with the option of using either the \defCCSD{} or \bogCCSD{} formulation.

Next, ground-state energies of nickel ($Z=28$) isotopes are displayed in Fig.~\ref{fig:nickelenergies} between $N=20$ (\elem{Ni}{48}) and $N=52$ (\elem{Ni}{80}). Closed-(sub)shell nuclei along this chain are \elem{Ni}{48,56,60,68,78}.
Similarly to the case of calcium isotopes, the CCSD results of the \magicint{} potential are closer to data than the \deltago{} potential.
As \deltago{} is a harder interaction than \magicint{}, it typically yields larger correlation energies at the CCSD level and triples contributions are therefore also expected to be larger. This is reflected in the gray band.
For both interactions the differences between \defCCSD{} and \bogCCSD{} reach $2-3$~MeV in $^{62,64}$Ni, consistent with the absence of a subshell closure at $N=34$, while discrepancies beyond $^{68}$Ni remain below $1$~MeV. This points to an accurate description of neutron-rich nickel nuclei.

\subsection{Detailed comparison around $N=30$}
\label{sec: detailed}
We turn to a detailed comparison of selected isotopes using the EOM and symmetry-broken CC variants.
We note that \elem{Ca}{50} and \elem{Ni}{58}  can be accessed via both 2PA-EOM and 2PR-EOM techniques based on the $N=28,32$ (sub-)shell closures at \elem{Ca}{48,52} and \elem{Ni}{56,60}, respectively.

\begin{figure}[t!]
    \centering
    \includegraphics[width=\columnwidth]{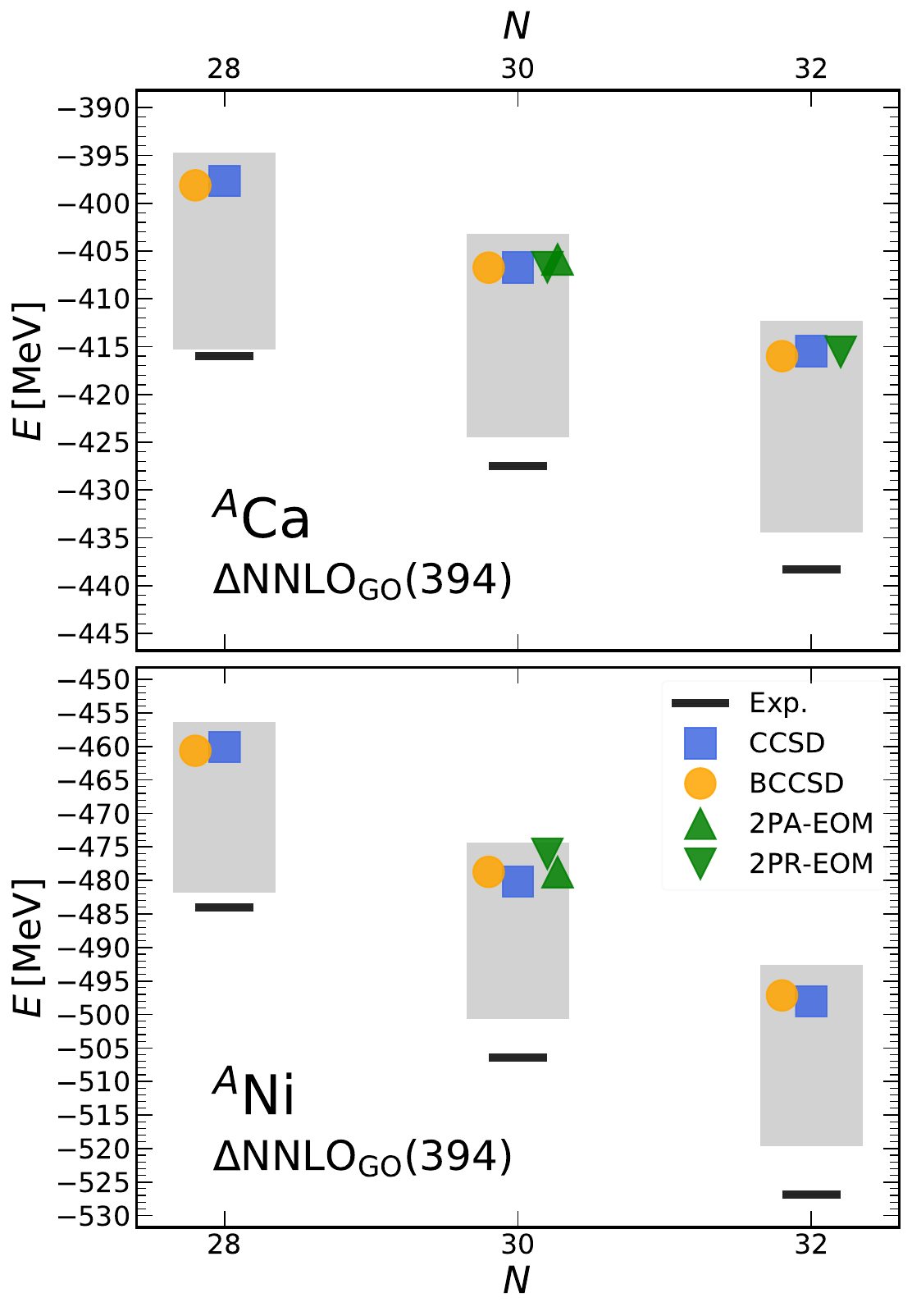}
    \caption{
    Ground-state energies for \elem{Ca}{48,50,52} (top) and \elem{Ni}{56,58,60} (bottom) isotopes as a function of the number of neutrons $N$.
    Calculations are performed with the \deltago{} interaction, and results are reported for \defCCSD{} (squares), \bogCCSD{} (circles), and, when available, 2PA-EOM (upward triangles) and 2PR-EOM (downward triangles).
    As in Fig.~\ref{fig:calciumenergies}, the shaded bands account for an estimate of the theoretical uncertainties due to the model-space truncation and the contribution of the missing triples.
    Experimental ground-state energies are shown as black bars.
    }
    \label{fig:detailed}
\end{figure}

Table~\ref{tab: detailed energies} and Fig.~\ref{fig:detailed} show ground-state energies of \elem{Ca}{48,50,52} and \elem{Ni}{56,58,60} obtained with the \deltago{} potential. 
\begin{table*}
    \caption{Ground-state energies of the Ca and Ni isotopes shown in Fig.~\ref{fig:detailed}. All calculations are performed using the \deltago{} interaction and are reported in MeV.
    Results for the \defCCSD{}, \bogCCSD{}, and (when available) 2PA-EOM and 2PR-EOM methods are shown.
    The last column  (labeled ``Triples'') reports expected triples contributions, estimated as 10\% of the \bogCCSD{} correlation energy. 
    }
    \label{tab: detailed energies}
    \begin{ruledtabular}
    \begin{tabular}{ p{1.9cm} p{2.7cm} p{2.7cm} p{2.7cm} p{2.7cm} p{2.7cm} }
                  & \defCCSD{} [MeV] & \bogCCSD{} [MeV] & 2PA-EOM [MeV]& 2PR-EOM [MeV]& Triples [MeV]\\
                  \hline\hline
    \elem{Ca}{48} & $-397.7$ & $-398.1$ &          &          & $-17.2$  \\
    \elem{Ca}{50} & $-406.7$ & $-406.7$ & $-405.9$ & $-406.7$ & $-17.8$  \\
    \elem{Ca}{52} & $-415.5$ & $-416.0$ &          & $-415.6$ & $-18.4$  \\
    \elem{Ni}{56} & $-460.1$ & $-460.6$ &          &          & $-21.2$  \\
    \elem{Ni}{58} & $-480.1$ & $-478.7$ & $-478.8$ & $-476.0$ & $-21.9$  \\
    \elem{Ni}{60} & $-498.0$ & $-497.2$ &          &          & $-22.6$  \\
    \end{tabular}
    \end{ruledtabular}
\end{table*}
In Fig.~\ref{fig:detailed} the shaded bands account for theoretical uncertainties. 
For convenience, only uncertainties on the BCC calculations are shown.
We can appreciate a good agreement in all the considered isotopes. The \bogCCSD{} and \defCCSD{} results differ by less than 1.5~MeV whereas
2PR-EOM and 2PA-EOM results are in near-perfect agreement in \elem{Ca}{50} with each other and the symmetry-breaking methods.
For \elem{Ni}{58}, 2PA-EOM, \defCCSD{}, and \bogCCSD{} results are close, too, while the 2PR-EOM results deviate slightly by 3~MeV.
This discrepancy may be related to \elem{Ni}{60}, which acts as a core for the 2PR-EOM computation of \elem{Ni}{58}, featuring a weak subshell-closure character~\cite{Otsuka2020}. This is supported by the spherical CCSD ground-state energy for \elem{Ni}{60} (not shown) being 5~MeV lower than the corresponding \defCCSD{} \acroGs{}~energy, while the two techniques typically agree within 1~MeV in closed-subshell isotopes.  
We note, however, that the discrepancies between the different techniques are within the estimated model-space uncertainties ($\approx 2-4\;\rm{MeV}$) and subleading with respect to the impact of the missing triples contributions (last column of Tab.~\ref{tab: detailed energies}).

\subsection{Two-neutron separation energies}

We focus on the two-neutron separation energies
\begin{align}
    S_{2n}(N,Z) \equiv E(N-2,Z) - E(N,Z) \, ,
\end{align}
which provide signatures of shell closures and serve as key indicators for identifying the neutron drip line.
Because neighboring isotopes are computed within the same Hamiltonian and many-body framework, their theoretical uncertainties are strongly correlated. Consequently, separation energies benefit from a substantial cancellation of systematic errors.
Corrections from neglected triples corrections are size-extensive and therefore scale linearly with particle number. Their contributions in neighboring nuclei are thus very similar, implying that their net effect on differential quantities such as $S_{2n}$ is strongly suppressed. A reliable description of two-neutron separation energies is therefore expected already at the present level of approximation.
Residual uncertainties may nevertheless remain, arising from the absence of symmetry restoration and, in neutron-rich isotopes close to the drip line, from missing couplings to the particle continuum~\cite{Hagen2012,Tichai2024}.

\begin{figure*}[ht!]
    \centering
    \includegraphics[width=\columnwidth]{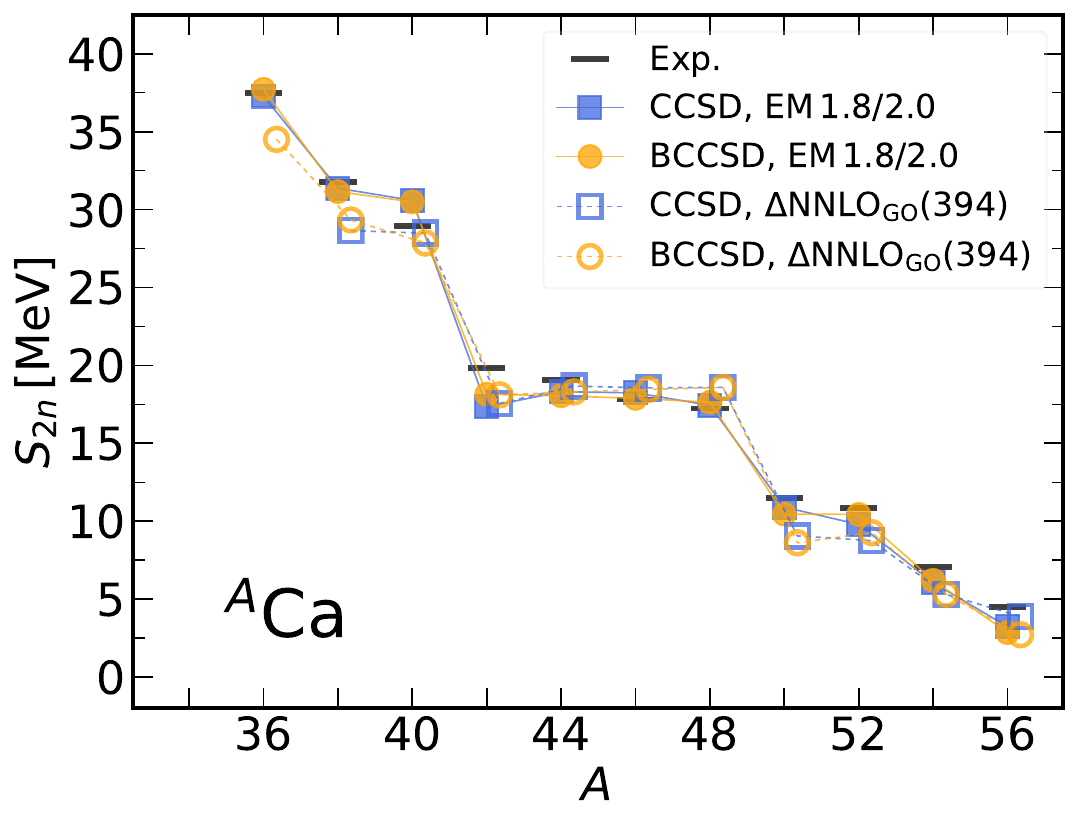}
    \includegraphics[width=\columnwidth]{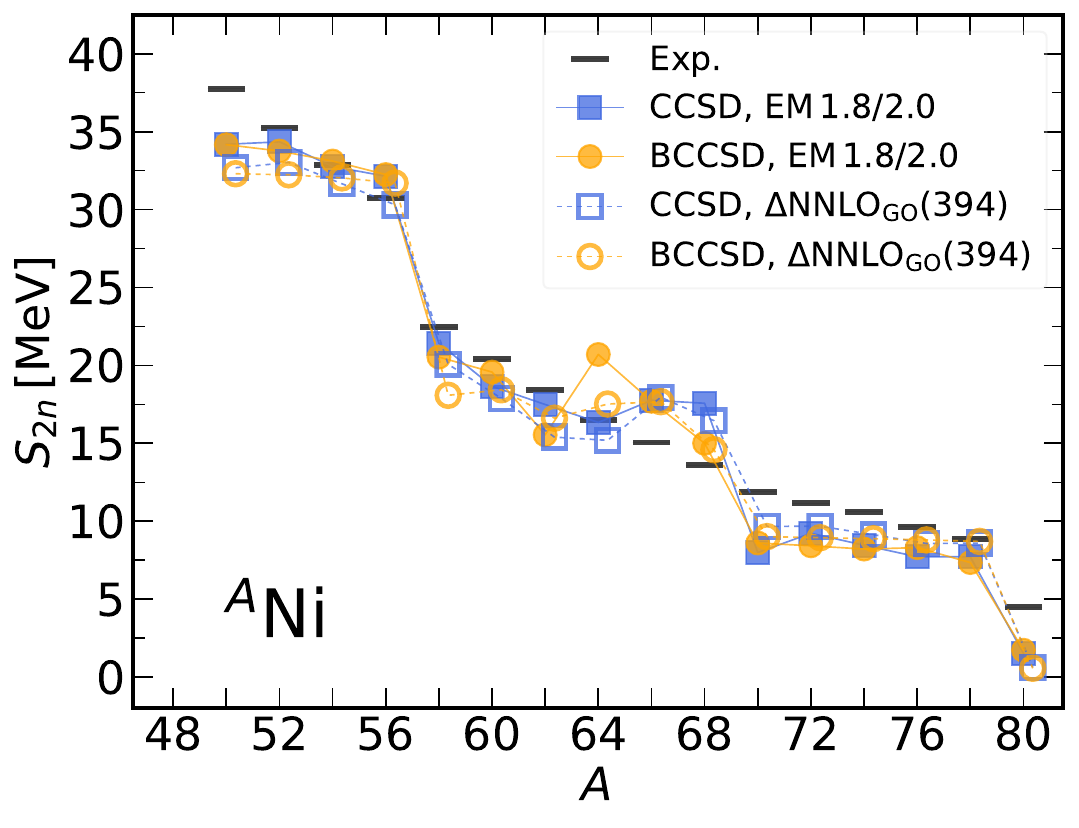}
    \caption{
    Two-neutron separation energies $S_{2n}(N,Z)$ as a function of the mass number $A$ in the Ca and Ni isotopic chains.
    Results obtained with the \magicint{} and \deltago{} potentials are shown as filled and empty symbols, respectively.
    Calculations performed with \defCCSD{} and \bogCCSD{} are denoted by squares and circles, respectively.
    Experimental data are shown as black bars.
    For clarity, the results for the \deltago{} interaction are offset horizontally by 0.3.
    }
    \label{fig:S2n}
\end{figure*}

Figure~\ref{fig:S2n} displays two-neutron separation energies along the calcium and nickel nuclei based on the \magicint{} and \deltago{} interactions. We focus on the symmetry-broken CC approaches, as these are the only methods that can systematically address all open-shell nuclei along the chains.

Let us first consider calcium isotopes. For both interactions, jumps are observed at \elem{Ca}{40} and \elem{Ca}{48}, accompanied by a nearly flat trend in between, \ie{} across the neutron $f_{7/2}$ shell.
Beyond \elem{Ca}{48}, another small plateau is reached as the $p_{3/2}$ shell is filled, followed by a weaker kink associated with the sub-shell closure at $N=32$. 
Deviations between \defCCSD{} and \bogCCSD{} are negligible and agreement with experiment is achieved for both potentials already at the singles and doubles truncation level. The largest discrepancies from the experimental data are about 2-3~MeV. 
Since many-body truncation effects are expected to largely cancel out for this observable, the dependence on the interaction model appears to be the leading source of uncertainty for $S_{2n}$ in the calcium chain. 

Let us now turn to the nickel isotopic chain. For both interactions and many-body methods, pronounced changes in slope appear at $N = 28$, $N = 40$, and $N = 50$, confirming the closed-shell character of $^{56}$Ni, $^{68}$Ni, and $^{78}$Ni, respectively. This is consistent with the relatively high $2^+_1$ excitation energy of these nuclei, exceeding $2$~MeV~\cite{nudat,Hagen2016,Taniuchi2019,SagawaHagino,Nowacki:2021fjw}. 
While \defCCSD{} and \bogCCSD{} closely track each other along the chain for the \deltago{} interaction, larger differences arise near $^{64}$Ni for the \magicint{} Hamiltonian.
Table~\ref{tab: detailed Nickel midshell} shows our results for this potential, \ie{} ground-state energies at both the mean-field and (B)CCSD levels, together with two-neutron separation energies and shell gaps computed with \defCCSD{} and \bogCCSD{}. 

Comparing calculations that break rotational symmetry (\defHF{}, \defCCSD{}) with those that break particle-number (\sphHFB{}, \bogCCSD{}), we find that \defHF{} and \defCCSD{} typically yield slightly more binding than their superfluid counterparts, except in the mid-shell isotopes \elem{Ni}{64,66}.
Indeed, \bogCCSD{} is 2~MeV more bound than \defCCSD{} in \elem{Ni}{64}, accounting for the fact that the $S_{2n}$ value in \elem{Ni}{64} computed with \bogCCSD{} exceeds the \defCCSD{} result by about 4.5~MeV.
At the mean-field level, these nuclei exhibit small yet non-negligible deformation ($\beta \approx$ 0.03).
The interplay between pairing and deformation in such mildly deformed systems can occasionally induce instabilities in the \sphHFB{} solutions, which subsequently propagate to the correlated \bogCCSD{} results.

\begin{table*}
    \caption{
    Ground-state energies, two-neutron separation energies $S_{2n}$, and two-neutron shell gaps $\Delta_{2n}$ obtained for Ni isotopes between $A=50$ and $A=78$ using the \magicint{} interaction.
    Ground-state energies are from the \defHF{}, \defCCSD{}, \sphHFB{}, and \bogCCSD{} methods.
    Separation energies and two-neutron shell gaps are from \defCCSD{} and \bogCCSD{}.
    All results are in MeV.
    }
    \label{tab: detailed Nickel midshell}
    \begin{ruledtabular}
    \begin{tabular}{ p{1.2cm} p{2.2cm} p{2.cm} p{2.cm} p{2.cm} | p{2.4cm} p{2.cm}  p{2.cm} p{2.cm} }
        & Deformed HF & CCSD & $S_{2n}$ & $\Delta_{2n}$ & Spherical HFB  & \bogCCSD{}  & $S_{2n}$ & $\Delta_{2n}$ \\
    \hline\hline
    \elem{Ni}{50} & $-262.8$ & $-374.8$ & 34.2 & $-0.2$ & $-261.9$ & $-374.9$ & 34.2 & 0.4 \\
    \elem{Ni}{52} & $-290.9$ & $-409.2$ & 34.4 & 1.6  & $-288.7$ & $-408.7$ & 33.8 & 0.6 \\
    \elem{Ni}{54} & $-317.6$ & $-441.9$ & 32.8 & 0.6  & $-316.1$ & $-441.8$ & 33.1 & 0.9 \\
    \elem{Ni}{56} & $-344.2$ & $-474.1$ & 32.1 & 10.7 & $-344.2$ & $-474.1$ & 32.3 & 11.7 \\
    \elem{Ni}{58} & $-361.4$ & $-495.5$ & 21.4 & 2.8  & $-358.8$ & $-494.6$ & 20.5 & 1.0  \\
    \elem{Ni}{60} & $-373.0$ & $-514.2$ & 18.7 & 1.2  & $-373.7$ & $-514.2$ & 19.6 & 4.0  \\
    \elem{Ni}{62} & $-387.9$ & $-531.6$ & 17.5 & 1.2  & $-389.3$ & $-529.7$ & 15.5 & $-5.2$ \\
    \elem{Ni}{64} & $-402.7$ & $-548.0$ & 16.3 & $-1.4$ & $-404.8$ & $-550.4$ & 20.7 & 3.0  \\
    \elem{Ni}{66} & $-418.0$ & $-565.7$ & 17.8 & 0.2  & $-414.0$ & $-568.1$ & 17.7 & 2.7  \\
    \elem{Ni}{68} & $-434.0$ & $-583.3$ & 17.6 & 9.6  & $-433.8$ & $-583.1$ & 15.0 & 6.4  \\
    \elem{Ni}{70} & $-438.4$ & $-591.3$ & 8.0  & $-1.2$ & $-438.1$ & $-591.7$ & 8.6  & 0.2 \\
    \elem{Ni}{72} & $-444.6$ & $-600.5$ & 9.2  & 0.8  & $-442.7$ & $-600.1$ & 8.4  & 0.2 \\
    \elem{Ni}{74} & $-449.8$ & $-609.0$ & 8.5  & 0.7  & $-447.7$ & $-608.3$ & 8.2  & $-0.1$ \\
    \elem{Ni}{76} & $-454.5$ & $-616.7$ & 7.7  & 0.0  & $-453.3$ & $-616.6$ & 8.3  & 1.0 \\
    \elem{Ni}{78} & $-459.5$ & $-624.4$ & 7.7  & 6.2  & $-459.1$ & $-624.0$ & 7.3  & 5.6 \\
    \end{tabular}
    \end{ruledtabular}
\end{table*}

Except for this particular case, $S_{2n}$ remains nearly flat between $^{58}$Ni and $^{68}$Ni, reflecting the gradual filling of the $p_{3/2}$, $f_{5/2}$, and $p_{1/2}$ neutron subshells.
Notably, and in contrast to the calcium isotopes, no signs of magicity are observed at $N = 32$ and $N = 34$ in the Ni chain~\cite{Steppenbeck2013,Otsuka2020}.
Finally, as already shown in Ref.~\cite{Tichai2024} for \magicint{}, the \deltago{} interaction also predicts $^{80}$Ni to be bound, suggesting that the drip line is located beyond $N=52$.

\subsection{Two-neutron shell gaps}
\label{sec: shell gaps}
A more refined understanding of the nuclear phenomenology can be obtained from two-neutron shell gaps, defined as (see Ref.~\cite{Scalesi:2024nao})
\begin{align}
    \label{eq:shell gap}
    \Delta_{2n} &\equiv S_{2n}(N,Z) - S_{2n}(N+2,Z) \\
    &= E(N-2,Z) - 2 E(N,Z) + E(N+2,Z). \nonumber
\end{align}
As $\Delta_{2n}$ is proportional to the second derivative of the ground-state energy at given neutron number, a sudden increase indicates a local minimum in the ground-state energy, and thus an additional gain in stability associated with a shell closure~\cite{Tichai2024,Scalesi:2024nao}.
This gives us a complementary way of probing shell effects without resorting to, \eg{}, computing excitation energies of $2_1^+$ states. 
 
\begin{figure*}
    \centering
    \includegraphics[width=\columnwidth]{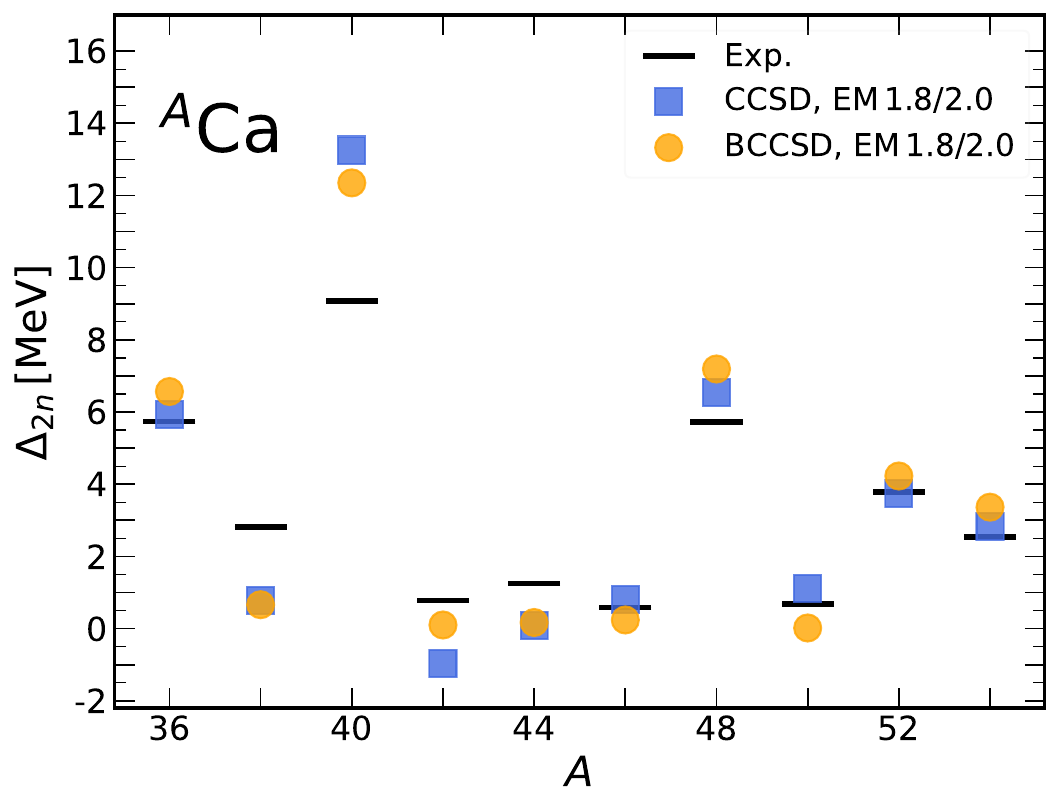}
    \includegraphics[width=\columnwidth]{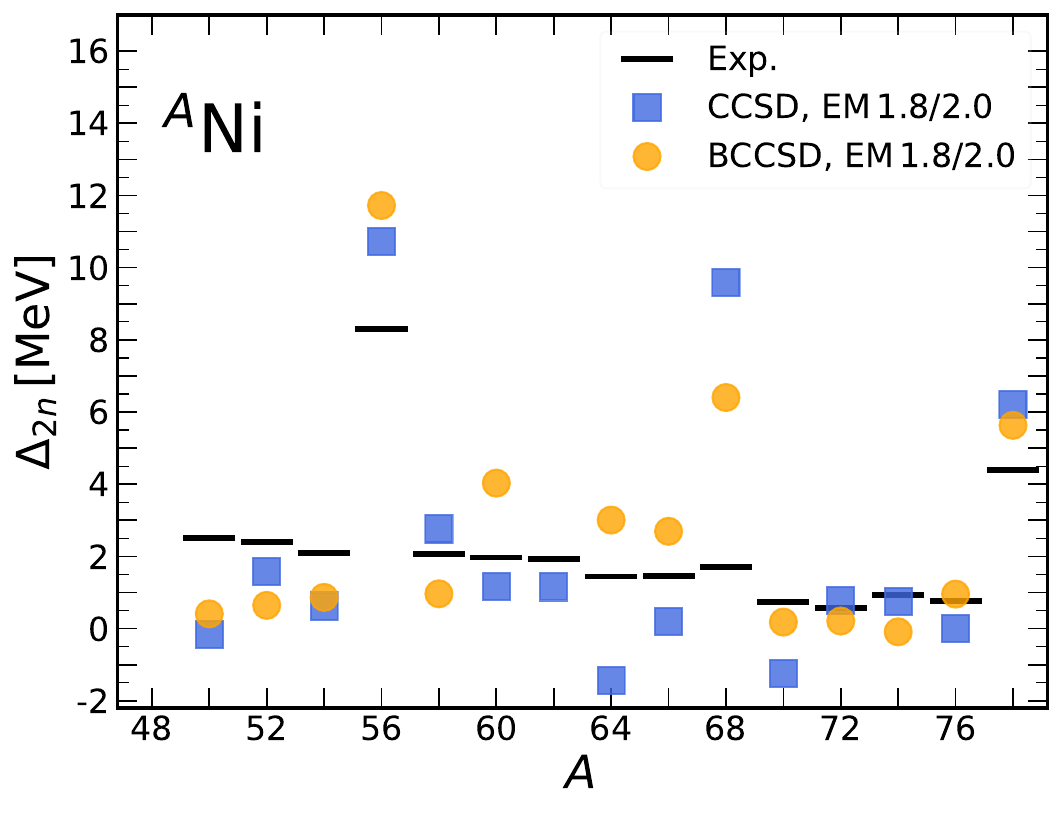}
    \caption{Two-neutron shell gaps in Ca (left panel) and Ni (right panel) isotopes computed with the \magicint{} interaction using the \bogCCSD{} and \defCCSD{} approaches.}
    \label{fig:gaps}
\end{figure*}

Figure~\ref{fig:gaps} shows the evolution of the empirical shell gaps along the Ca (left) and Ni (right) isotopic chains. We used the \defCCSD{} and \bogCCSD{} approaches and the \magicint{} interaction. We note that the \deltago{} potential yields a similar picture.
As a benchmark, we show numerical data for nickel in Table~\ref{tab: detailed Nickel midshell}. 

For calcium isotopes between \elem{Ca}{36} and \elem{Ca}{54}, both symmetry-broken approaches agree.
A strong enhancement of $\Delta_{2n}$ is evident in \elem{Ca}{36,40,48} in conjunction with $N=16$~\cite{Miller2019,Lalanne2023}, $N=20$, and $N=28$ shell closures.
While smaller in absolute terms (approximately 4~MeV), the two-neutron shell gaps in the neutron-rich \elem{Ca}{52,54} magic nuclei~\cite{Steppenbeck2013,Otsuka2020} stand out compared to the neighboring \elem{Ca}{50} or the average value (about 2~MeV) seen for open-shell Ca nuclei with $42 \le A \le 46$.

For nickel, shown in the right panel of Fig.~\ref{fig:gaps}, $\Delta_{2n}$ features two strong maxima at $N=28$ (\elem{Ni}{56}) and $N=50$ (\elem{Ni}{78}). This is consistent with the corresponding kinks observed in the two-neutron separation energy. Again,  results from \defCCSD{} and \bogCCSD{} are close to each other, in particular for \elem{Ni}{78}. Both approaches, however, overpredict the experimental shell gap at $N=28$ by about $30\%$. 

Interestingly, the enhancement of the shell gap for \elem{Ni}{68}, already predicted by \bogCCSD{} in Ref.~\cite{Tichai2024}, is qualitatively confirmed by \defCCSD{} calculations, although the discrepancy is somewhat large in this case (about 3~MeV). 
However, the computations predict magicity at $N=40$ while this is not observed in the experimental two-neutron shell gaps. It seems that the CC computations at the singles and doubles truncation level tend to enhance the closed-shell character of \elem{Ni}{68} in both $S_{2n}$ and $\Delta_{2n}$.
The agreement between \defCCSD{} and \bogCCSD{}, in general, is rather good for $A\le 56$ and $A\ge70$.
Deviations turn out to be larger for open-shell isotopes lying between $N=28$ and $N=40$. Two-neutron shell gaps tend to enhance the discrepancies noticed for separation energies.

\section{Conclusion and outlook}
\label{sec: Conclusions}

In this work, we compared three different coupled-cluster approaches to ground states of open-shell nuclei.
We presented an overview of the two-particle-attached and two-particle-removed EOM-CC methods and discussed the impact of adopting a symmetry-breaking reference state in the single-reference CC expansion.
Our calculations focused on the even semi-magic Ca and Ni nuclei, allowing us to compare the ground-state energies and two-neutron separation energies predicted by EOM-CC (for nuclei close to subshell closures), Bogoliubov CC, which includes pairing correlations in the reference state, and CC based on a deformed Slater determinant.

We found that ground-state energies computed with the different CCSD methods agree within uncertainty estimates from model-space truncations. Furthermore, differences are always significantly smaller than the estimated contribution of triples excitations. 
A substantial agreement between CC variants is also observed for the two-neutron separation energies. Moreover, being a differential quantity, for which systematic errors cancel, separation energies are in accordance with the experiment already at the singles and doubles truncation level, even in very neutron-rich isotopes.

The consistency of the different CC implementations underscores the strength of CC theory as a tool for describing nuclear phenomenology in mid-mass systems, with the potential to scale at a reasonable computational cost to very heavy open-shell nuclei.
The symmetry-broken variants investigated here offer clear advantages over particle-attached and particle-removed EOM formulations, as their scope extends to mid-shell isotopes which are beyond the reach of low-rank excitation operators in the EOM ansatz.
This distinction becomes particularly important in heavier systems, where shell sizes increase substantially and only nuclei in the vicinity of major shell closures can be targeted by EOM-CC.
In addition, symmetry-broken reference states can be tailored to the intrinsic structure of the target nucleus: deformed reference states are well suited for strongly collective, doubly open-shell systems, whereas superfluid spherical reference states are appropriate for heavy semi-magic chains.

Ultimately, a unified treatment that incorporates both deformation and pairing correlations within a single many-body framework remains desirable. While proof-of-principle studies have demonstrated the promise of such an approach (see Ref.~\cite{Signoracci2015}), its large-scale implementation remains to be explored.

\begin{acknowledgements}
We thank W.~Jiang and U.~Vernik for helpful discussions.
This work was supported in part by the Deutsche  Forschungsgemeinschaft  (DFG,  German Research Foundation) through Project ID 279384907 -- SFB 1245  and through Project ID 39083149 -- PRISMA${}^+$ EXC 2118/1, by the European Research Council (ERC) under the European Union's Horizon Europe research and innovation programme (Grant Agreement No.~101162059), by Research Foundation Flanders (FWO, Belgium, grant 11G5123B), by the U.S. Department of Energy, Office of Science, Office of Nuclear Physics, under award No.~DE-FG02-96ER40963 and the FRIB Theory Alliance award DE-SC0013617; 
by the U.S. Department of Energy, Office of Science, Office of Advanced Scientific Computing Research and Office of Nuclear Physics, Scientific Discovery through Advanced Computing (SciDAC) program (SciDAC-5 NUCLEI); 
by the Alexander von~Humboldt Foundation.
This research used resources of the Oak Ridge Leadership Computing Facility located at Oak Ridge National Laboratory, which is supported by the Office of Science of the Department of Energy under contract No. DE-AC05-00OR22725. 
The authors gratefully acknowledge the Gauss Centre for Supercomputing e.V. (www.gauss-centre.eu) for funding this project by providing computing time through the John von Neumann Institute for Computing (NIC) on the GCS Supercomputer JUWELS at J\"ulich Supercomputing Centre (JSC). Computer time was also provided by the Innovative and Novel Computational Impact on Theory and Experiment (INCITE) program and the supercomputer Mogon at Johannes Gutenberg Universit\"at Mainz.
\end{acknowledgements}

\section*{DATA AVAILABILITY}
The data that support the findings of this article are
openly available~\cite{marino_2026_18655344}.

\bibliography{bibliography}

\end{document}